# Project Lx Conventos: Travelling through space and time in Lisbon's religious buildings

João Gouveia, Fernando Branco, Armanda Rodrigues and Nuno Correia

CITI, Faculdade de Ciências e Tecnologia, Universidade Nova de Lisboa, Lisbon, Portugal

**Abstract**

Project Lx Conventos aims to study, in a systematic and integrated manner, the impact of the dissolution of religious orders in the dynamics of urban transformation in nineteenth century Lisbon. After the liberal revolution and the civil war, in the 19th century, the dissolution of religious orders led to the alienation, in Lisbon, of nearly 130 religious buildings which were then given profane occupations (mainly public services) or demolished and divided in plots, originating new urban realities. Project Lx Conventos thus aims to show that the extinction of the convents was decisive in the urban development of Lisbon, in the eighteen hundreds.

The project stands on a large set of multimedia data which includes historic and contemporary cartography and geo-referenced photos, videos and 3D models, provided by the projects partners, Lisbon Municipality and the Portuguese National Archive, Torre do Tombo.

Supported by these materials, the project's team is creating an online system that will implement a spatial and temporal navigation of these resources integrated in an interactive Map of Lisbon. Besides spatially locating and analyzing the data available for each of the religious buildings considered in the project, the tool integrates cutting edge interaction technology for: 1) Enabling a temporal voyage over the available traces of religious buildings; 2) Analyzing the evolution of religious buildings and their surroundings, through available data; 3) Using 3D representations of the buildings for accessing related data, through time. In this paper, the tools under development in the context of Lx Conventos are described, as well as the technologies supporting them. The current status of the system is presented and future developments are proposed.

# 1. Introduction

The goal of the Lx Conventos project is to study the impact of the dissolution of religious orders in the dynamics of urban transformation in nineteenth century Lisbon, focusing on the historical evolution of the different religious houses located in the city. In this context, an online interactive platform was developed enabling users and researchers to visualize and study both the buildings' and their surroundings' temporal change.

The project stands on a large set of multimedia data which includes historic and contemporary cartography and geo-referenced photos, videos and 3D models, provided by the projects partners, The city of Lisbon Municipality and the Portuguese National Archive, Torre do Tombo. The initial requirements for the system involved its online availability, based on a cartographic interface, laying out the geographic mapping of the project's focus of study: the religious buildings and their surroundings. Moreover, several types of searches should be made available, as well as the integration of different interaction techniques, enabling a temporal voyage over the available traces of religious buildings.

On a more technical note, the developed platform has been prepared to receive different types of data originating from various time periods and to build a temporally evolutive structure of these data. This structure is then used to generate an interactive interface for visualizing the temporal evolution of buildings under study. Both the system interface and storage structures are, thus, independent of the types and dimensions of available data for each building. As each building under study is unique on its own, the information available, over the years, portraying its evolution, may vary in terms of time, dimension, media and format. The system must, thus, be able to generate the proper structures and needed functionality to provide the user with the right tools to visualize and interact with the available resources, consistently and transparently. Also, the set of tools developed, and mainly the Image Matching Tool, described in section 4, are independent from the types of data available and from the aim of the project's platform itself. These tools focus on showing evolution of existing structures in space. This evolution could be temporal, but other alternatives are possible.

In this paper, we present the current state of the platform, which is divided in two parts: the first one consists mainly of project dissemination information. This area includes general and case studies, glossary, news and legislation, as well as a textual query interface for the religious houses; The second part of the platform consists of a temporal and spatial navigation interface, which begins on a map representation of Lisbon. On this map, the user can visualize the spatial representation of the different religious buildings, in different timelines. By choosing one particular building, the user accesses its information and may further navigate on its 3D evolutive representation. This 3D model addresses the temporal evolution of the building with the help of photos taken in different time periods. In order to create this complex visualization, a separate tool is being developed, to help structure the information in such a way that the platform can later on read it and construct the visualization.

This article is structured in the following manner: in this section, we introduce the Lx Conventos project and its objectives. In section 2 the most important related work for reaching the proposed goals is described. In section 3 the proposed project platform, including the map application functions and the associated types of temporal visualizations, is presented.

After this section, the Image Matching Tool functionalities are described and how it is used to create the interactive visualization. Finally, the paper ends with the conclusions and the description of the future work for this project.

## 2. Related work

Due to recent advancements in image creation/manipulation technology, a large amount of images are being produced on a daily basis. This situation created interest in developing different and creative techniques to handle large amounts of image data. Specifically, in this paper, we concentrate on systems and techniques enabling the transmission of historical knowledge through the identification of spatial and temporal relationships in sets of dated images.

Basically, the relevant research threads involve 3D model navigational tools and different interactive techniques, which focus on the manual alignment of images [1]. Moreover, the concept of Rephotography [2] was also analysed and considered, due to the requirement to enable the use and connection of data originating from different periods of time.

The following four related systems were analysed: **Photo Tourism** [3], **Photosynth**[1], the system described in **Computational Rephotography** [4] and **Third View**[2]. The last two systems are both implementations of the Rephotography concept.

Photo Tourism is a browsing system which provides a navigational interface for a 3D model, consisting of a large collection of images [3]. The 3D visualization is a representation of the scenery, to which views representing the perspective used for capturing the photograph are added as a form of links (Figure 1a). When a view is selected, a close-up image representing it, along with other relevant information, like a short description, are presented to the user. Simultaneously, the system shows a sequence of thumbnails related to the previous submitted query and a panel with photos related to the current one (Figure 1b).

Photosynth is based on Photo Tourism. The system receives a collection of images, finds their common points and creates a 3D model composed of these images. With this model, users can navigate between related images as if it were a full 3D environment.

Rephotography [2] is a concept that is intended to combine images from the past and present in order to provide a spatio-temporal evolution visualization. The system [4] uses a camera and a computer to reproduce a reliable rephotography of a certain image. A computer processes the discrepancies between the former and the current image and provides feedback to move the camera, capturing an image with the same distance and perspective as the original.

The Third View web application also uses the concept of Rephotography. However, instead of using a camera to produce a reliable image, it merges existing photos of the same location, taken with the same perspective, providing the viewer with the possibility of visualizing the differences between different periods in time.

---

[1] Photosynth - http://photosynth.net/ last access in July 2014
[2] Third View - http://www.thirdview.org/3v/rephotos/index.html last access in July 2014

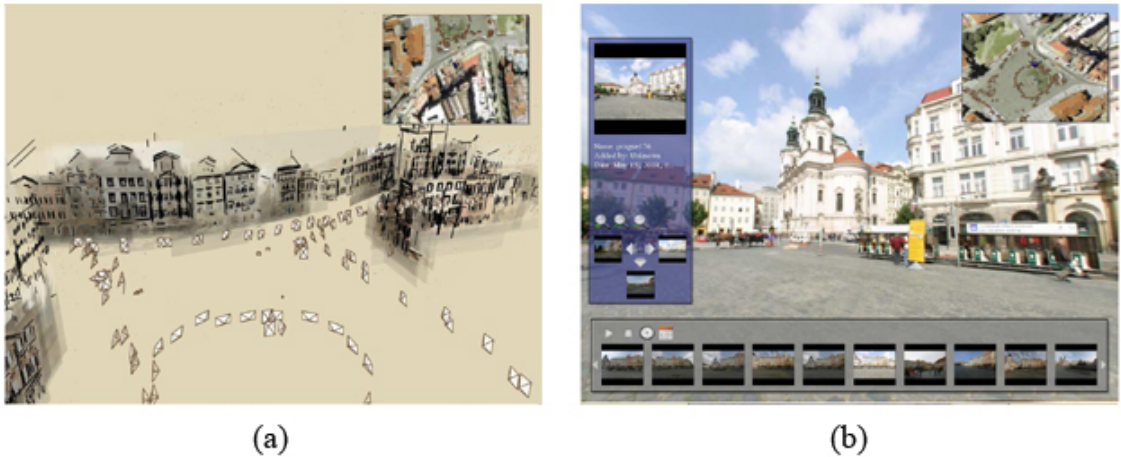

*Figure 1 - Images from Photo Tourism [3]. (a) 3D visualization of the scenery, with views from the captured photographs. (b) Tool state after selecting a view from image represented in (a).*

Additionally, some more commercial, already user proven, products were also found to be relevant to our studies. Out of those, Streetmuseum[3] is important to highlight for this research. This is a mobile application that uses different types of data, like pictures, text and a 3D view to teach its users about the history of London. The application shows relevant information as the user walks around the city and points its camera to relevant buildings. The important places are marked on a map.

Besides the systems described, it is also important, for this research, to study relevant image processing algorithms. The most significant was SIFT [5], which searches and extracts interest points from an image and then generates descriptions of the features near the points, called descriptors. These descriptors are essential to find the same features in different images and then to create relationships between them.

Since some structures or geographic regions have completely changed over the years, it may be hard to automatically find similar objects in images from different time periods, and thus use existing procedures to relate the images. Because of this, interactive techniques are needed to generate links between images [1]. These techniques allow the user to manually select common points between images which are later used to create geometric relationships.

## 3. Lx Conventos System - The evolution of Lisbon's religious houses

The developed Lx Conventos Platform can be divided in two different parts. The first one consists of a data dissemination module, which endows users with access to project resources and deliverables. The second part of the platform consists of an interactive navigational tool, based upon an online map of Lisbon.

*3.1. Lx Conventos Platform*

The platform is composed of four main sections: 1) an about page, where the project is described; 2) a dissemination section which will publish the findings of the different project studies; 3) a glossary, legislation and news pages; and (4) a search section, where users may search for the different religious

---

[3] http://www.museumoflondon.org.uk/Resources/app/you-are-here-app/home.html last access in July 2014

houses located in Lisbon. The search may be initiated in two different ways: a text-based search and a map-based one. This latter method corresponds to the second part of the platform and is explained below. A prototype image of the website front page can be seen on Figure 2.

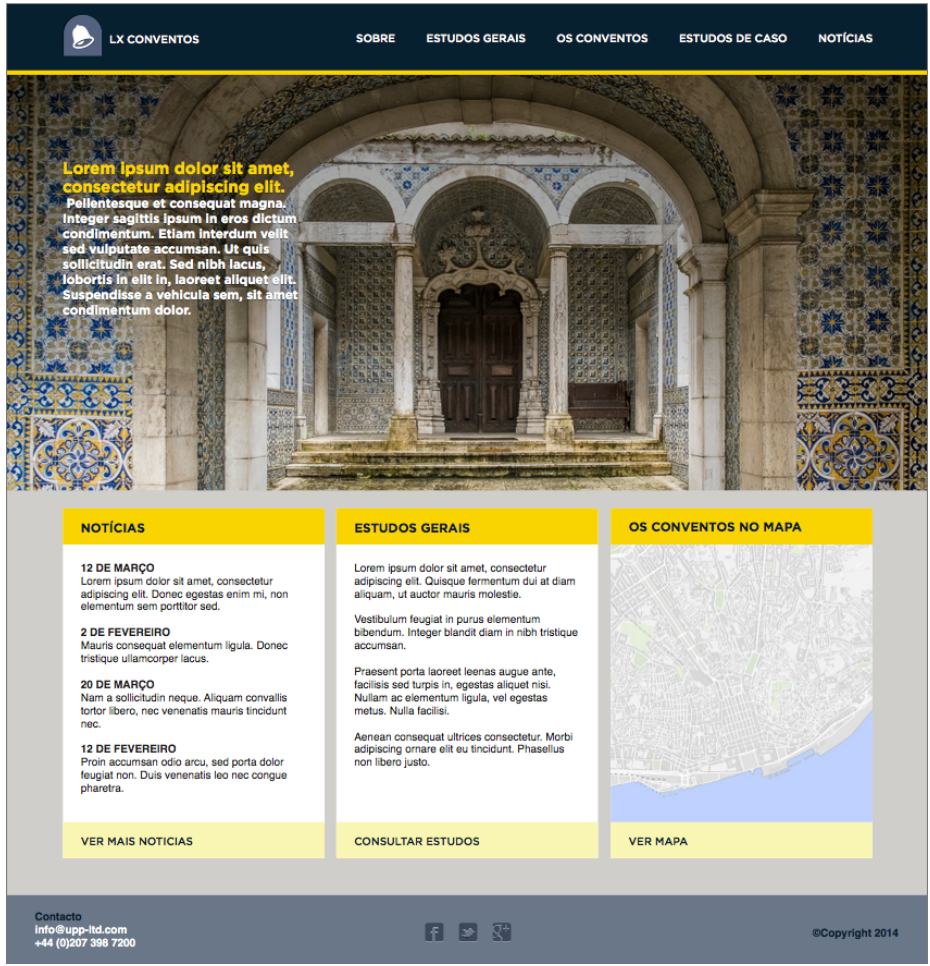

*Figure 2 - Proposed design of Lx Conventos front page.*

*3.2. Map navigational tool*

The main goal of this platform is to show users a temporal and spatial evolution of Lisbon religious houses, without previous knowledge of the available resources for each specific building. In order to achieve this, two different visualizations were created: the first consists of a navigational interface on a map-based tool, enabling access to the geographically placed buildings and surroundings; the second one enables interaction with the existing resources on a specific religious house, by creating and making available a representation model constructed from those resources.

The map shows the location and spatial mapping of Lisbon religious houses in 1834 and in the present day (in 2014). This helps the user visualize their spatial evolution: Did they lose structure? Was the garden shape altered? Did they disappear entirely? Were new avenues built on top of them? This interface can answer these types of questions. Moreover, the map interface enables the overlaying, with different levels of transparency, of historic cartographic maps of Lisbon, giving the users another level of comparison, between the old maps and the layers of religious houses. On Figure 3 we can see the overlapping of an old cartographic map with the current map. The religious house overlays are also visible. In this case, the cartographic map has 50% of transparency.

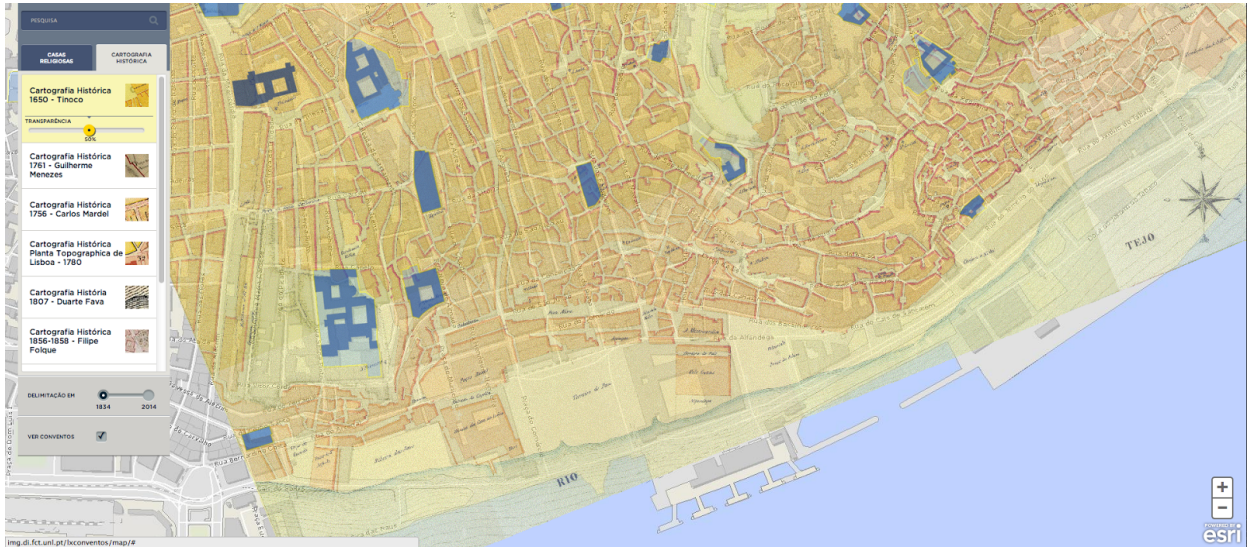

*Figure 3 - An old cartographic map, with 50% transparency, on top of the current map of Lisbon. The religious houses overlay is shown in blue, presenting the buildings as they were in 1834.*

When a user clicks on a religious house (Figure 4) a window pops up, showing an image of the selected house, as well as a short description. Further action enables a "visit" to the building, opening a window where related information is shown. As seen in Figure 5, this information may include descriptive text, a gallery of photos, a gallery of drawings or a 3D visualization. This visualization will include, as a basis, a 3D model of the house. This model acts as a support for the navigation and, on top of it, based on a selected timeline date (Figure 6), different pictures are shown. This enables the visualization of the evolution of the building through time. Since the available photos may originate from different time periods, it is also possible for the user to see the differences on the surrounding background. The user can select different timelines and directly see the differences in the relevant photos. Since this is a 3D representation, it is possible to navigate through rotation and see different sides of the building and detail, as long as photos are available. This visualization is shown in Figure 7.

To support the creation of this visualization another tool is being developed. This tool allows to create links between images which cannot be automatically related, when changes in the scenery are very complex. These connections will then be added to the 3D model of the building. Although the tool requirements were mainly derived from the project needs, its development is generic and it may be used in other projects, to improve 3D representations involving evolution of the objects being shown. In section 4 we describe this tool in more detail.

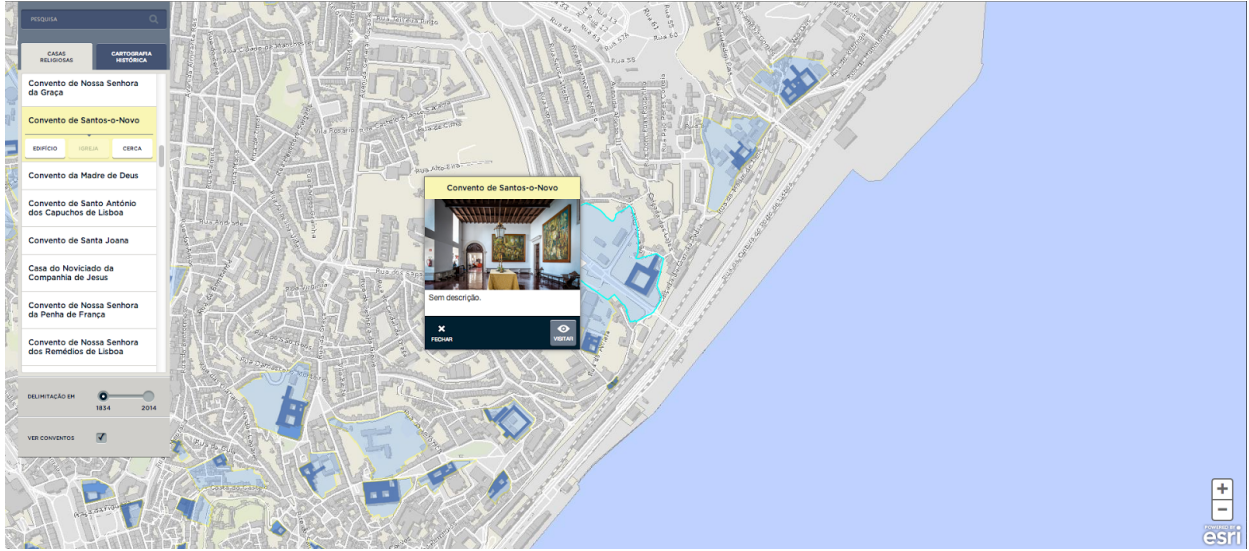

*Figure 4 - The selection of a building opens a window that shows a thumbnail picture and short description.*

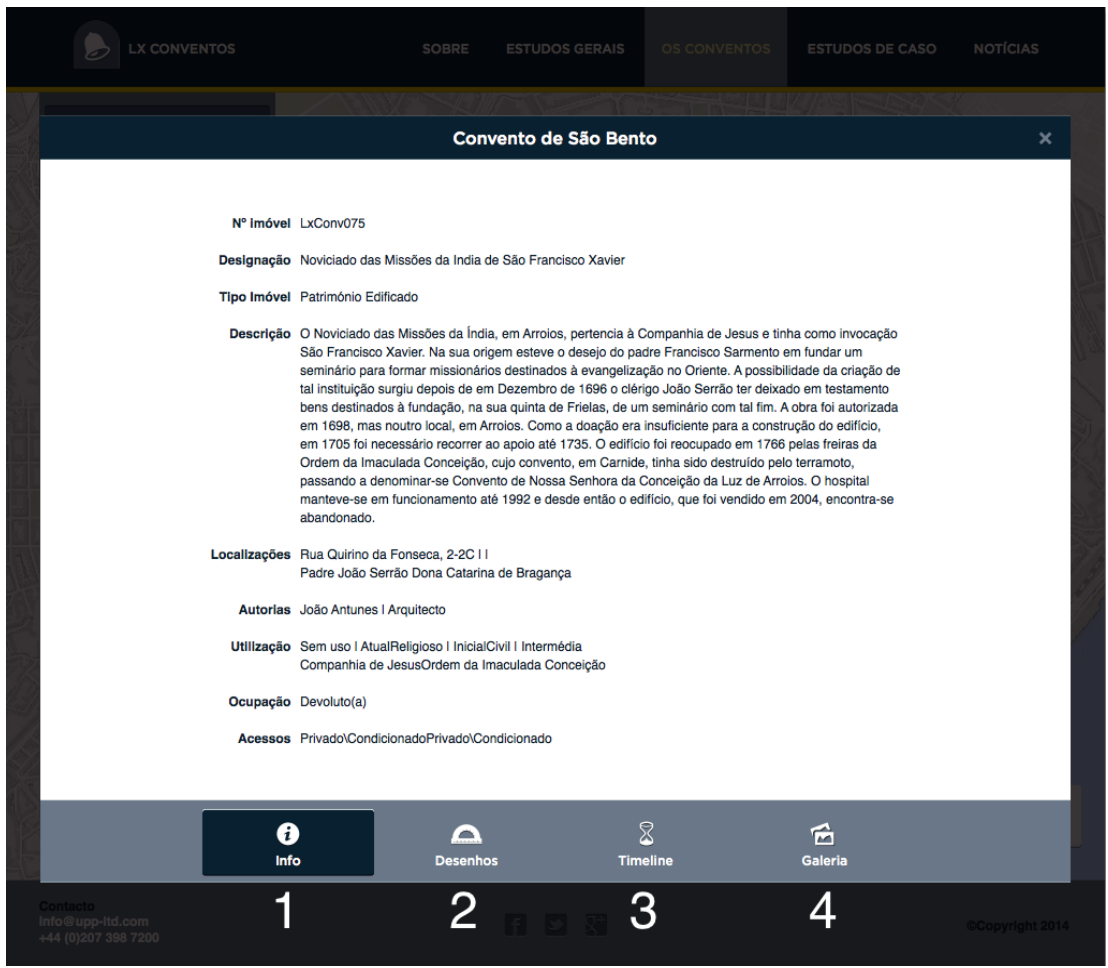

*Figure 5 - 1. Textual information; 2. Drawing gallery; 3. 3D Visualization; 4. Photo gallery.*

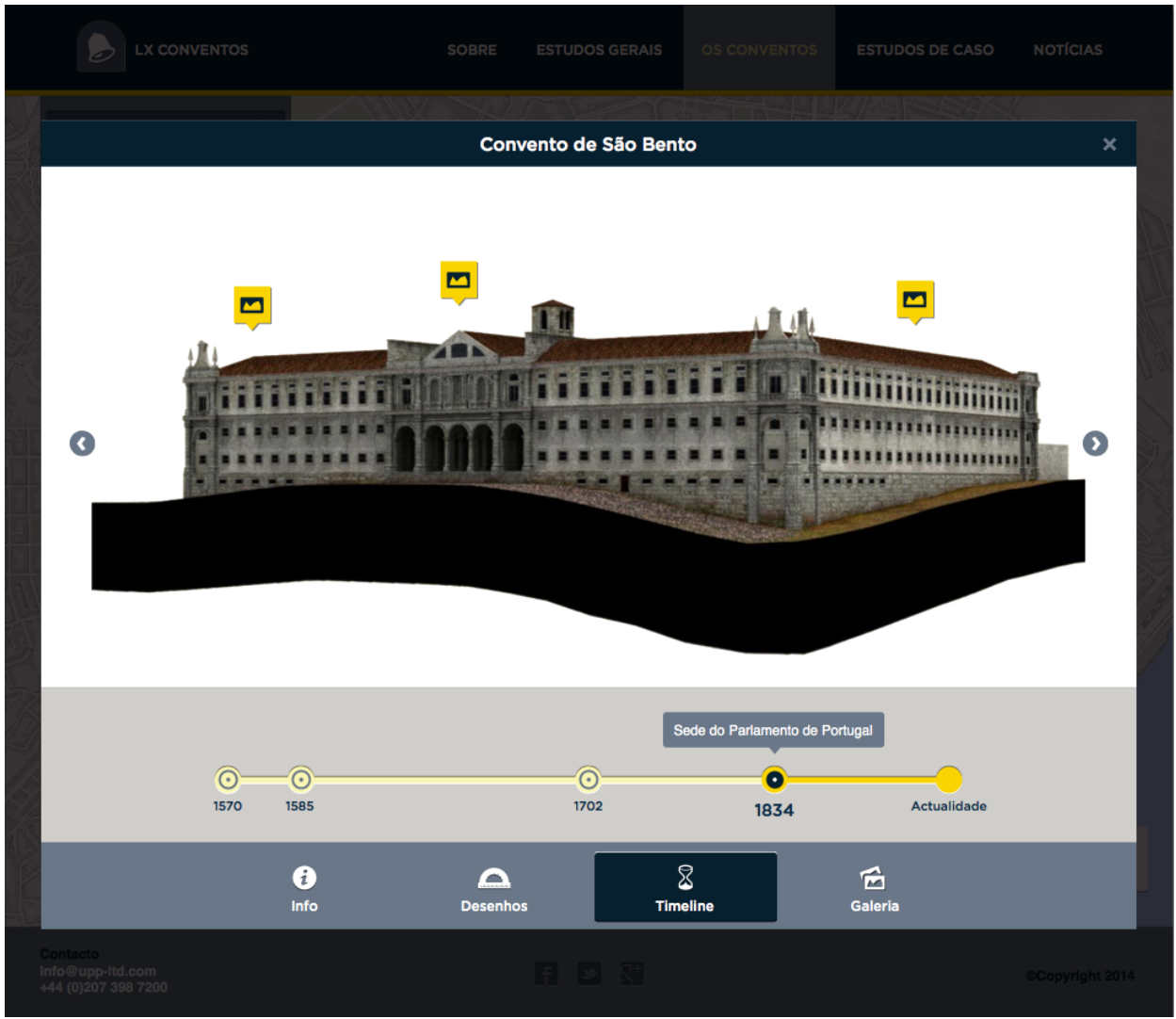

*Figure 6 - The timeline shows relevant historic dates and selecting one of them activates locations on the 3D model where content is available.*

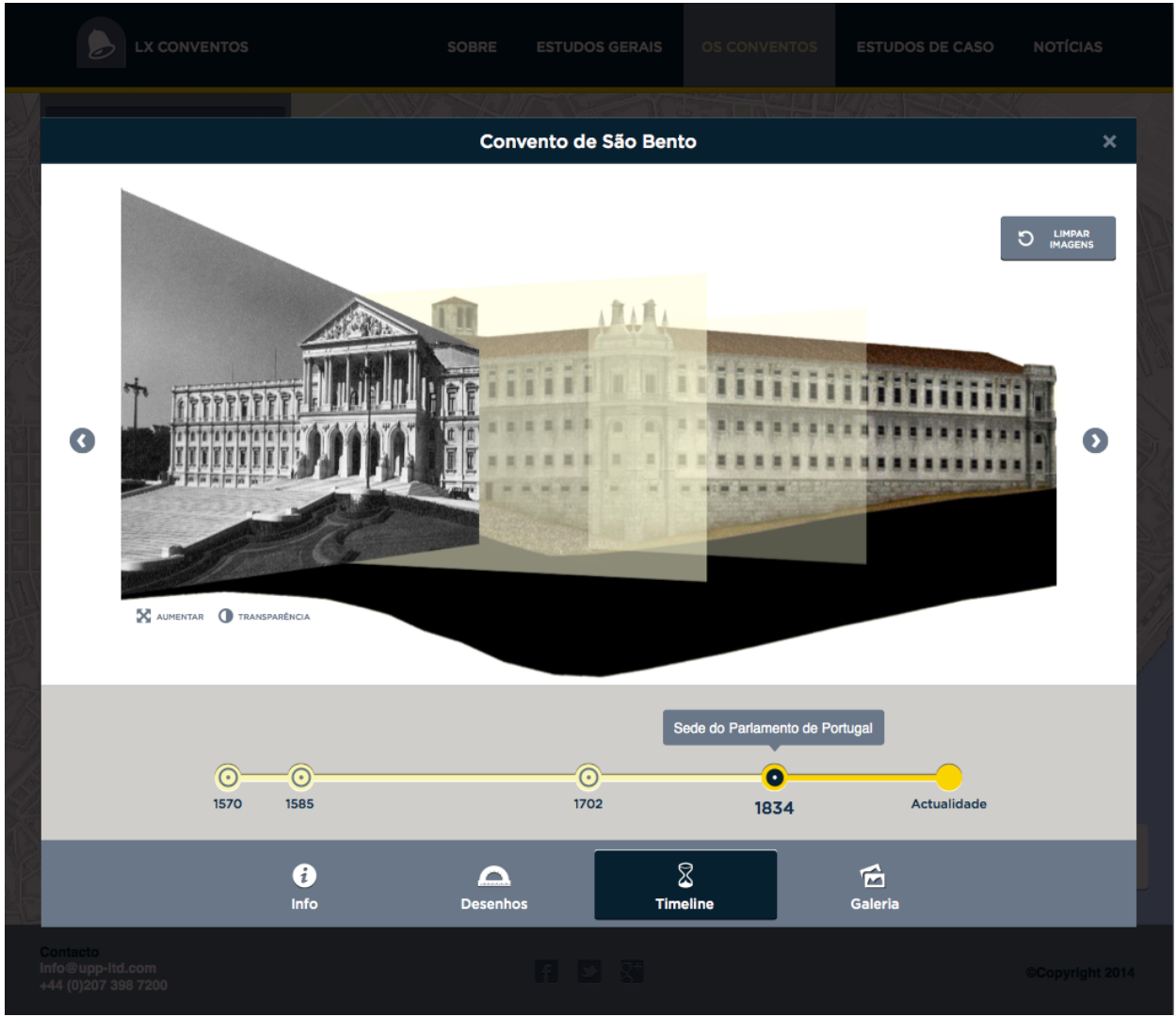

*Figure 7 - Pictures of a specific period overlayed on the 3D model. This model can be rotated to show a different perspective.*

## 4. Image Matching Tool

One of the aims of the Lx Conventos System is to show the spatio-temporal evolution of Lisbon religious houses, using the available images. One way to do this is to enable a temporally evolutive navigation, which may be automatically generated from temporally ordered images, when these include sufficiently similar details. As this is not the case in many of the resources available in this project, we are currently developing a tool, which enables to identify regions of similarity in different images. These regions will then be used to create links between the images, thus supporting the enrichment of the 3D model of the religious house. The Image Matching Tool is being developed in C++ with openFrameworks, as the main framework, and OpenCV, as the foundation for the developed image processing algorithms.

The major functionalities of the tool involve:

1. The automatic grouping of images from a complete set of pictures;

2. The interactive identification of regions of similarity between images, when these were not identified automatically;

3. The creation of links between similar images;

4. The import and export of these connections from/to a specifically formatted file, which will be used by the Lx Conventos Platform to create the model of the religious house.

When the tool is opened the user has the option to upload a set of images. After all images are loaded, it is possible to automatically generate groups. This functionality will use image processing algorithms, such as SIFT [5], to find similar characteristics between all loaded images (1). When the algorithm finds at least one similar characteristic between two or more images, it groups them together. On Figure 8 it is possible to see the groups of images the algorithm generated. Sometimes the automatic generation of groups may provide false positives, or the user may want to group images where the algorithm is unable to detect similarities. For example, if a building has endured many changes over time, two photos of it, of distinct moments, may be very different. When this happens the user can manually group pictures, two at a time, which enables functionality 2), interactively identifying regions of similarity between images. This is done by creating a four point polygon on both pictures, that corresponds to an area where the similarity should exist. Figure 9 shows this functionality, where a user identified and marked a region of similarity on both images. After this step, a connection between these two images is created (functionality 3). In Figure 10, an integrated view of several related images, using the perspective of the central one, is presented. This visualization is focused on the central image, distorting the remaining ones.

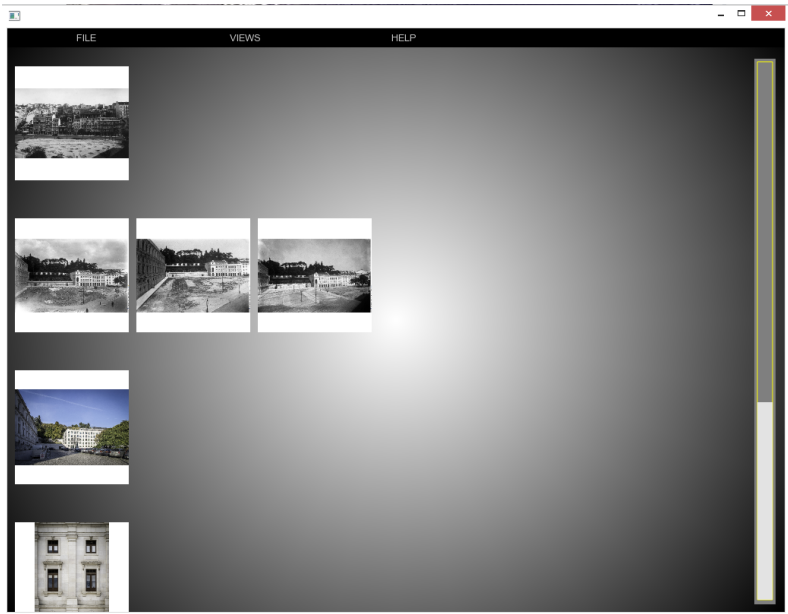

*Figure 8 - Automatically created groups. Each group represents images that are related to at least another image of the same group.*

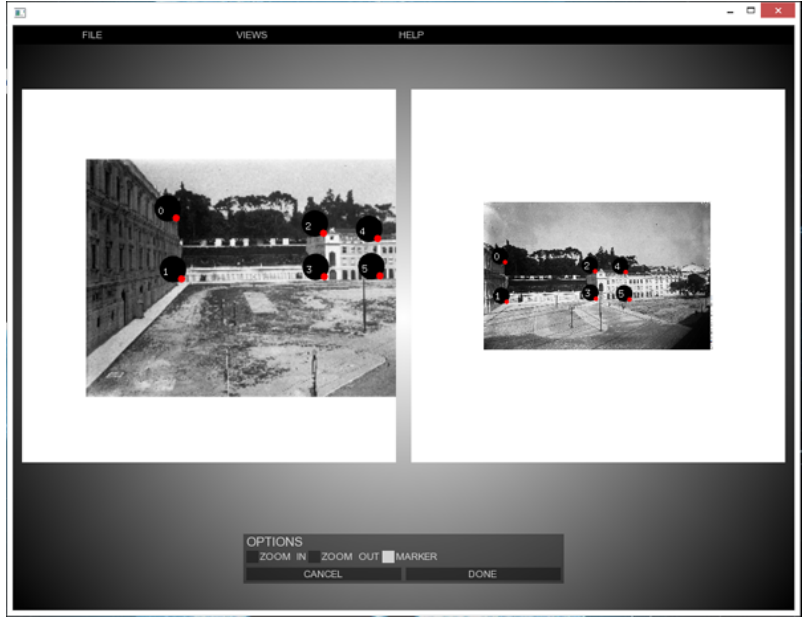

*Figure 9 - Manual selection of similar points between two images to create a relationship between them.*

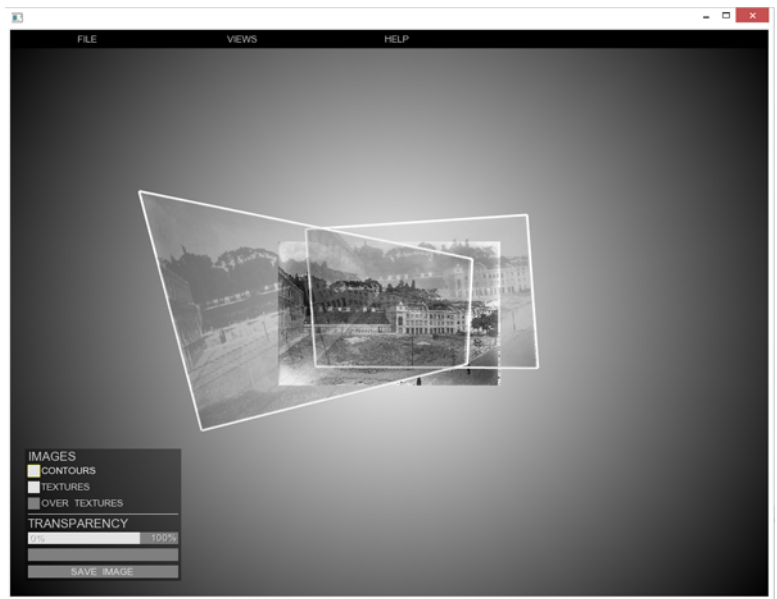

*Figure 10 - Visualization of an image and its relationships with other images in the same group.*

All links created between images can be exported and imported (functionality 4). The file exporting process is needed to enable the use of the generated information by other platforms (in this specific case, the Lx Conventos platform). The export functionality will facilitate the use of these visualizations, such as the one on Figure 10, both on the Lx Conventos platform, as well as, on other future platforms. The tool also imports the same file format, to enable storing and changing of several stages of the work.

## 5. Conclusions and Future Work

The initial aim of this project was the integration of the spatial mapping of Lisbon's religious houses with existing historic maps. Moreover, several related media were available, particularly from various time periods. One of the main points of the research was to show the temporal evolution of these buildings as they have existed for a long time and have been through many changes. The main visualization where the user could visualize both the temporal changes of a building and its surroundings through a metaphor of a visit, proved to be a good approach. In order to achieve this, a separate tool to match the different images together was required. The development of this separate tool also helped the creation of an application which can later be used to create other types of temporal-based visualizations. This expands our tool for other types of uses in other scientific areas.

It is important to make sure that both the tool and the online platform are on par for the use by different types of users. The next step is to further test the system in order to assess both the functionality and the user interface design. This will help evaluate the full potential of the created tools and detect where further development should be concentrated. When this is done we aim to find other areas of study where we can apply the image matching tool and help create different visualizations.

## Acknowledgements

The work described in this publication was supported by "Fundação para a Ciência e a Tecnologia", Portugal, through project Lx Conventos, ref. PTDC/CPC-HAT/4703/2012. The authors would like to thank the support provided by the team of Lx Conventos in Câmara Municipal de Lisboa, as well as the data used in the development of the Lx Conventos Platform. The contribution of Bárbara Teixeira in web design is also acknowledged.